\documentclass{INTERSPEECH2023}

\interspeechcameraready

\title{Adapting Language-Audio Models as Few-Shot Audio Learners}
\name{
Jinhua Liang$^{\star}$,
Xubo Liu$^{\ast}$,
Haohe Liu$^{\ast}$,
Huy Phan$^{\dagger}$,
Emmanouil Benetos$^{\star\ddagger}$,\\
\it Mark D. Plumbley$^{\ast}$,
Wenwu Wang$^{\ast}$
}
\address{
$^{\star}$ Centre for Digital Music, Queen Mary University of London, UK \\ 
$^{\ast}$ Centre for Vision, Speech and Signal Processing (CVSSP) \\
$^{\dagger}$ Amazon Alexa 
$^{\ddagger}$ The Alan Turing Institute, UK
}
\email{\{jinhua.liang, emmanouil.benetos\}@qmul.ac.uk, huypq@amazon.co.uk, \\ 
\{xubo.liu, haohe.liu, m.plumbley, w.wang\}@surrey.ac.uk}

\begin{document}

\maketitle
 
\begin{abstract}
Contrastive language-audio pretraining (CLAP) has become a new paradigm to learn audio concepts with audio-text pairs. CLAP models have shown unprecedented performance as zero-shot classifiers on downstream tasks. To further adapt CLAP with domain-specific knowledge, a popular method is to finetune its audio encoder with available labelled examples. However, this is challenging in low-shot scenarios, as the amount of annotations is limited compared to the model size. In this work, we introduce a \textbf{Tr}aining-\textbf{eff}icient (Treff) adapter to rapidly learn with a small set of examples while maintaining the capacity for zero-shot classification. First, we propose a cross-attention linear model (CALM) to map a set of labelled examples and test audio to test labels. Second, we find initialising CALM as a cosine measurement improves our Treff adapter even without training. The Treff adapter beats metric-based methods in few-shot settings and yields competitive results to fully-supervised methods.
\end{abstract}

\section{Introduction}
\label{sec:intro}
Learning new concepts from a small set of examples is challenging in machine learning. It is a de-facto issue in audio domains where high-quality labels are more labor-intensive to obtain. While existing few-shot algorithms exploit available annotations by directly learning in the few-shot setting (i.e., $n$-way $k$-shot problem)~\cite{vinyals_matching_2016,snell_prototypical_2017}, contrastive language-audio pretraining (CLAP) provides a new paradigm to learn audio concepts using large-scale audio-text pairs. CLAP has shown an impressive capacity for zero-shot knowledge transfer in audio classification~\cite{guzhov_audioclip_2022,elizalde_clap_2022,wu_large-scale_2023}, largely due to 
the use of a large-scale datasets and its different training objective.

To further adapt CLAP to downstream datasets, existing methods directly finetune the audio encoder with examples in target domains~\cite{elizalde_clap_2022,wu_large-scale_2023}. However, finetuning is ill-suited for few-shot settings because the number of available examples is extremely small with respect to the number of model parameters. In addition, updating the parameters of CLAP's audio encoder with discriminative learning would break the connection to its language encoder. A natural question is: \textit{can we bootstrap a CLAP in few-shot scenarios while maintaining its capacity as a zero-shot classifier?}

\begin{figure}[t]
    \centering
    \includegraphics[scale=0.8]{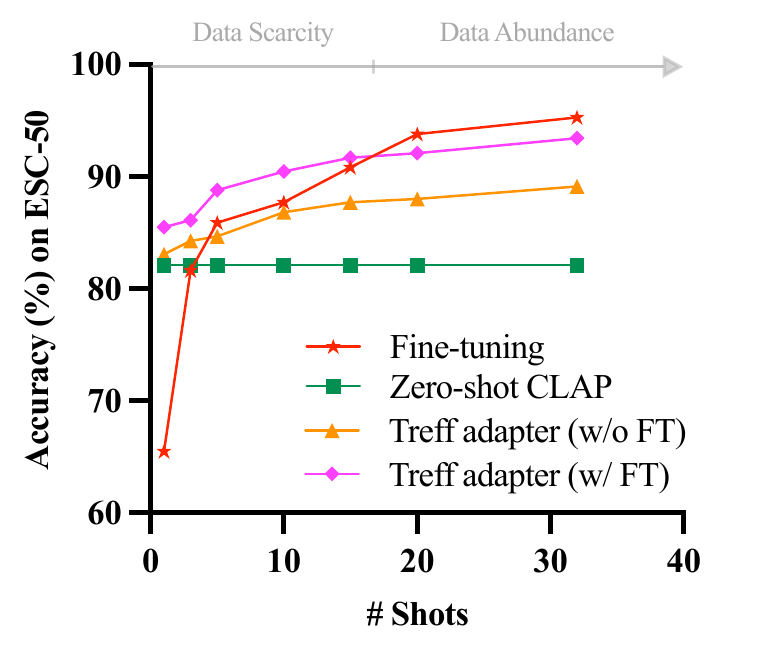}
    \caption{\small{\textbf{Adaptation accuracy on the ESC-50 dataset v.s. numbers of shots across various models}. Our Treff adapters outperform other adaptation approaches in data-scarce scenarios and are comparable to finetuning with abundant data. We find that current approaches to domain adaptation yield a poor performance in low-shot settings. Finetuning is even worse than zero-shot knowledge transfer when fewer than 3 examples are available, largely due to overfitting. This motivates our work on few-shot domain adaptation.}}
    \label{fig:num_shots_increase}
\end{figure}

In this paper, we propose the Training-efficient (Treff) adapter to bootstrap CLAP models by bridging zero-shot knowledge transfer and few-shot finetuning. Our Treff adapter contains two core designs: a cross-attention linear model (CALM) and cosine initialisation. First, we devise CALM where a set of support audio clips and test audio clips are mapped to their probability distributions. Second, we find that with a proper initialisation, using CALM as a cosine similarity measurement yields a better performance even without training. Figure~\ref{fig:num_shots_increase} summarises 
performance of various adaptation approaches on the ESC-50 dataset v.s. the number of examples (i.e., shots) per class. Our Treff adapters outperform other adaptation approaches in few-shot scenarios when the number of examples per classes is less than 16. As the number of shots increases, the proposed Treff adapter can still have a competitive result to the finetuning approach. We find that our empirical study is consistent with some previous findings on different datasets~\cite{wang_who_2021}. Our experimental results demonstrate that the proposed Treff adapter outperforms zero-shot classification by a large margin with only a small set of annotated data. In addition, our Treff adapter finetuned with half the number of shots can have a comparable performance with state-of-the-art fully-supervised learners. We speculate this work would shed light on domain adaptation in data-limited scenarios. The contributions in this work are summarised below:

\begin{figure*}[th]
    \centering
    \includegraphics[scale=0.7]{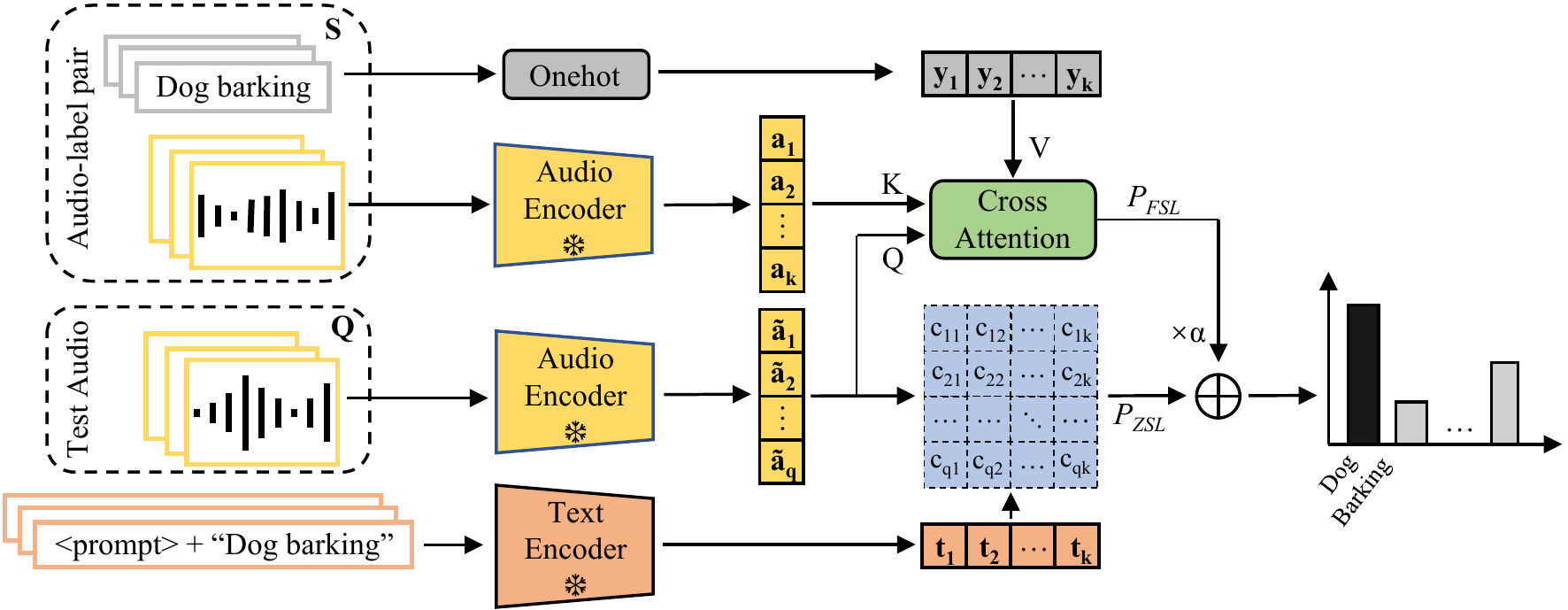}
    \caption{\small{\textbf{Overall framework of Treff adapter} to classify $q$ audio clips in an $n$-way $k$-shot setting. For the few-shot classification results  $P_{FSL}$, support embeddings $\mathbf{a}$ and test embeddings $\mathbf{\tilde{a}}$ are extracted with a frozen audio encoder separately. Test embeddings $ \mathbf{\tilde{a}}$ then attend to support embeddings $\mathbf{a}$ and their corresponding one-hot labels $\mathbf{y}$ in the CALM. To get the zero-shot transferred knowledge $P_{ZSL}$, candidate labels concatenated with a fixed prompt (e.g., ``This is a sound of ") are encoded by the frozen language encoder to get text embeddings $\mathbf{t}$. Coefficient $\{ \mathbf{c}_{qk}\}$ is then measured by calculating cosine similarity between test audio embedding $\mathbf{\tilde{a}}_q$ and text embedding $\mathbf{t}_k$. The overall result is yielded by combining $P_{FSL}$ and $P_{ZSL}$ with a learnable parameter $\alpha$.}}
    \label{fig: treff_adapter}
\end{figure*}

\begin{itemize}[itemsep=-2pt, topsep=0pt, leftmargin=0.5cm]
    \item We introduce the Treff adapter to further improve CLAP performance with a small set of examples while preserving its ability to zero-shot classification. To our knowledge, this is the first work to study audio-language models in the few-shot audio domain.
    \item We propose CALM to retrieve labels of the test examples by measuring the affinity between test and support embeddings. We show that the CALM can efficiently learn from few-shot settings while maintaining the semantic information of the pretrained feature representation.
    \item We devise a cosine initialisation strategy for CALM. We find that our Treff adapter benefits from this strategy in the few-shot settings even without training. The proposed Treff adapter in training-free version outperforms the zero-shot classifier by over 5\% accuracy on the ESC-50 dataset.
\end{itemize}

\section{Related work}
\label{sec:related_work}
\textbf{Contrastive cross-modality pretraining.} CLAP~\cite{elizalde_clap_2022,wu_large-scale_2023} follows the idea of Contrastive Image-Language Pretraining (CLIP)~\cite{radford_learning_2021} where contrastive learning is applied to train a cross-modal retrieval model with a large number of image-text pairs. The cross-modal retrieval model is trained so that multi-modal embeddings from the same pair are moved closer, otherwise, pushed away. Therefore, it can classify audio clips by measuring the similarity between multi-modal embeddings without additional examples (referred as zero-shot learning, ZSL).
Such cross-modality models, however, cannot benefit from extra labelled data during inference.  To further adapt cross-modality retrieval models with available examples, learnable prompts are devised to reduce CLIP's reliance on prompt engineering~\cite{zhou_learning_2022,zhou_conditional_2022}. Still, these adaptation methods require a large amount of labelled data for finetuning. The training-free adapter for CLIP (TIP-adapter) was proposed to learn a cache model to map test audio to their labels and initialise the cache model with support audio embeddings to avoid training~\cite{zhang_tip-adapter_2022}. It initialised the weights of the cache model with the values of support embeddings for rapid learning. The TIP-adapter, however, inevitably jeopardises the semantic information of support examples in the gradient-based optimisation. Cross-modal few-shot learning is designed where the examples from other modalities are treated as additional examples~\cite{lin_multimodality_2023}. Although this cross-modal method showed promising results in few-shot classification, it requires updating linear classifiers and multiple encoders for the best performance, which takes additional computational resources.

\noindent \textbf{Metric-based few-shot learning.} Few-shot learning (FSL) aims to train a generic encoder via $n$-way $k$-shot problems where a classifier to predict the probability distribution of test data (queries) over $n$ different classes and each class has $k$ labelled objects (support shots or shots) for reference. Let $(\mathbf{x}_i, \mathbf{y}_i)\in \mathcal{D}$ be a data point in the dataset $\mathcal{D}$. Suppose $\mathcal{S}=\{(\mathbf{x}_i, \mathbf{y}_i)\}_{i=1}^{nk}$ be a (small) support set of $nk$ examples. Given a test audio clip $\mathbf{\tilde{x}}_q\in \mathcal{Q}$ where $\mathcal{Q}$ is the test set, a few-shot paradigm is expected to predict categorical probabilities $P(\mathbf{\tilde{y}}_q)$: 
\begin{equation}
\label{eq:fsl}
    P(\mathbf{\tilde{y}}_q)=\mathcal{M}_{\theta}(\mathbf{\tilde{y}}_q|\mathbf{\tilde{x}}_q, \mathcal{S})
\end{equation}
where $\mathcal{M_\theta}$ is any mapping function and $\theta$ represents the model parameters. Metric-based FSL aims to cluster data points as per their categories in the emebedding space. Matching Networks map a set of support data and queries to their labels~\cite{vinyals_matching_2016}. Prototypical networks learn prototypes to represent different classes and measure the $l_2$ distance between queries and prototypes in the embedding space~\cite{snell_prototypical_2017}. Subsequent works attempted to apply prototypical networks to multi-label cases~\cite{liang_leveraging_2022}. Siamese networks take two different data as input and learn the distance between them explicitly~\cite{koch_siamese_2015}. 

\section{Treff adapter for CLAP}
\label{sec:proposed_method}
\subsection{Overall framework}
\label{subsec: overall_framework}
We hereby introduce the Treff adapter, a simple adaptation method to boost audio classification performance of the CLAP model by leveraging few-shot learning. 

As shown in Fig. \ref{fig: treff_adapter}, the Treff adapter makes use of audio-label pairs and zero-shot knowledge by aggregating logits of FSL and ZSL. For FSL, the Treff adapter extracts audio embeddings $\mathbf{A}=[\mathbf{a}_1, \mathbf{a}_2,\dots,\mathbf{a}_{nk}]\in \mathbb{R}^{nk\times d}$ and text embeddings $\mathbf{T}=[\mathbf{t}_1, \mathbf{t}_2,\dots, \mathbf{t}_n]\in \mathbb{R}^{n\times d}$ where $\mathbf{a}$ and $\mathbf{t}$ are column vectors of $\mathbf{A}$ and $\mathbf{T}$, respectively, and $d$ is the dimension of the embedding in the shared audio-language space. The test audio embedding $\tilde{\mathbf{A}}=[\mathbf{\tilde{a}}_1, \mathbf{\tilde{a}}_2,\dots,\mathbf{\tilde{a}}_{q}]\in \mathbb{R}^{q\times d}$ is likewise extracted by the audio encoder and then mapped to its label with the support audio-label pairs using CALM. The resulted affinity matrix $\mathbf{S}=\{s_{ij} \}$ is then combined with one-hot labels of support audio for few-shot classification. In the other branch, the cosine distances between the query audio embedding $\mathbf{\tilde{A}}$ and the text embedding $\mathbf{T}$ are measured and utilised for zero-shot classification explicitly. The overall inference for test audio $\mathbf{\tilde{y}}_q$ aggregates logits from zero-shot learning ($P_{ZSL}$) and few-shot learning ($P_{FSL}$):
\begin{equation}
    \label{eqn: overall_logits}
    P_{overall}(\mathbf{\tilde{y}}_q)=P_{ZSL}(\mathbf{\tilde{y}}_q|\mathbf{\tilde{x}}_q, \mathbf{T})+\alpha \cdot P_{FSL}(\mathbf{\tilde{y}}_q|\mathbf{\tilde{x}}_q, \mathcal{S})
\end{equation}
where $\alpha$ is a trainable parameter and we initialise it with 1.0 in our experiments. 

Our Treff adapter has two core designs: CALM and cosine initialisation. We illustrate how the CALM structure retrieves the label of a query audio clip using a small amount of support audio examples in Section~\ref{subsec: calm}; and in Section~\ref{subsec: training_efficient_adaption}, we show that our Treff adapter can function as a similarity measurement without finetuning by adopting a simple initialisation.

\vspace{-0.2cm}
\subsection{CALM}
\label{subsec: calm}
\begin{figure}[th]
    \centering
    \includegraphics[scale=0.7]{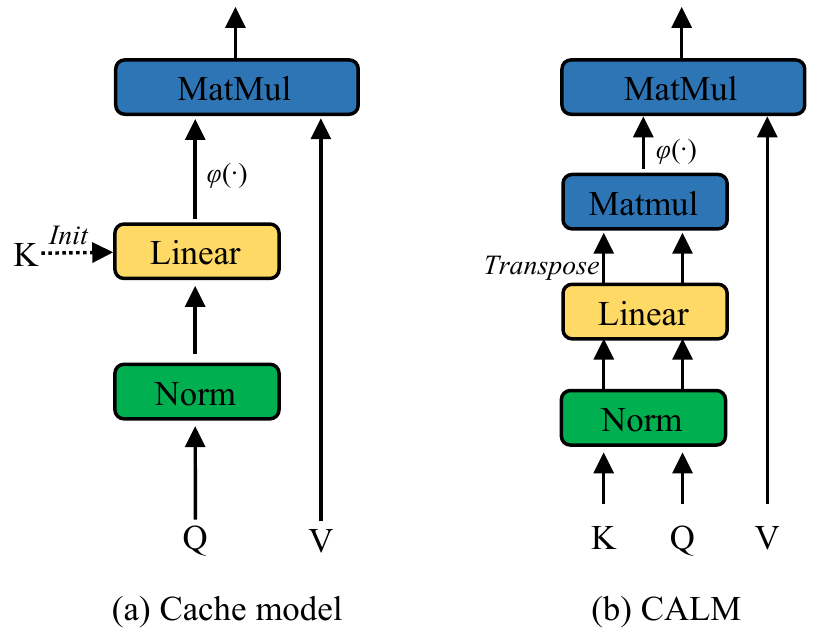}
    \caption{\small{\textbf{Cache model (left) and our proposed CALM (right)}. The $Q$ and $K$ values are the query and support audio embeddings respectively, while the $V$ values are the one-hot labels associated to support audio examples.}}
    \label{fig: xattention}
\end{figure}

Inspired by recent progress in attention mechanisms~\cite{vaswani_attention_2017}, we devise CALM by using the $KQV$-attention~\cite{vaswani_attention_2017}. Different from common cross-attention modules where $K$ and $V$ are from the same modality~\cite{jaegle_perceiver_2022,alayrac_flamingo_2022}, CALM uses bi-modality information from labelled examples. Figure~\ref{fig: xattention} compares structures of the cache model~\cite{zhang_tip-adapter_2022} and our CALM where $Q$ and $K$ denote the query and support audio embeddings, respectively, and $V$ denotes the one-hot labels of the support examples. The query embedding $\mathbf{\tilde{a}}_i$ and the support embedding $\mathbf{a}_j$ are normalised and then fed into a trainable linear layer, separately
\begin{equation}
    \label{eqn: transform_query_embedding}
    \mathbf{\tilde{e}}_i=f_{\mathbf{W}}(\mathbf{\tilde{a}}_i)=\frac{\mathbf{\tilde{a}}_i}{|\mathbf{\tilde{a}}_i|} \mathbf{W}^\mathsf{T},
\end{equation}
\begin{equation}
    \label{eqn: transform_support_embedding}
    \mathbf{e}_j=f_{\mathbf{W}}(\mathbf{a}_j)=\frac{\mathbf{\tilde{a}}_j}{|\mathbf{a}_j|} \mathbf{W}^\mathsf{T},
\end{equation}
where $\mathbf{W}$ is the weight in the trainable layer. We note that the linear layer shares the weights across these two embeddings. The motivation is that sharing the weights reduces the trainable parameters in the finetuning. The linear layer maps the audio embedding to a new space where the embeddings from the same class get closer to each other. 
The affinity coefficient is then calculated by
\begin{equation}
    \label{eqn: affinity_matrix}
    s_{ij}=\mathbf{\tilde{e}}_i^\mathsf{T}\mathbf{e}_j,
\end{equation}
The output of the CALM is obtained by
\begin{equation}
    \label{eqn: few_shot_logits}
    \mathbf{o}_i=\sum_j\varphi(s_{ij})\cdot y_j,
\end{equation}
where $\mathbf{o}_i\in \mathbb{R}^n $ is the probability distribution over the $i$-th test audio example;~$\varphi(\cdot)$ is a scaling function controlling the sharpness of the similarity coefficient
\begin{equation}
    \varphi(x)=\exp(b(1-x)) 
\end{equation}
where $b$ is a temperature factor, set empirically as 5.5 in our experiments.

\vspace{-0.2cm}
\subsection{Training-efficient adaptation}
\label{subsec: training_efficient_adaption}
\textbf{Training-free Treff adapter.} The Treff adapter endows CLAP models with the ability to learn new, domain-specific knowledge from a small set of labelled examples. We prove that the Treff adapter can boost the model performance in a few-shot setting even \textit{without} gradient-based optimisation. Inspired by the TIP-adapter~\cite{zhang_tip-adapter_2022}, a generic encoder pretrained on a large-scale database should be good enough to cluster data points as per their semantic information. We replace the weights of the trainable linear layer with an identity matrix when no fine-tuning is available. Therefore, coefficients $s_{ij}$ in the affinity matrix would be
\begin{equation}
    \label{eqn: affinity_matrix_without_finetune}
    s_{ij}=\biggl(\frac{\mathbf{\tilde{a}}_i}{|\mathbf{\tilde{a}}_i|}\biggl)^\mathsf{T}\frac{\mathbf{a}_j}{|\mathbf{a}_j|},
\end{equation}

\noindent In this case, the non-trainable version of CALM is actually a cosine-similarity function between the query and support embedding (i.e., $\cos(\mathbf{\tilde{a}}_i, \mathbf{a}_j)$). We found that even the cosine-similarity scores between the audio embeddings are capable of indicating how close those audio examples are.

\noindent \textbf{Finetuning the Treff adapter.} While the training-free Treff adapter outperforms the performance of zero-shot knowledge transfer, it is still not comparable with finetuning and other gradient-optimization methods (e.g., the TIP-adapter~\cite{zhang_tip-adapter_2022}). We thus optimise the weights of the Treff adapter with the support examples only. Specifically, we initialise the weights of trainable linear layer with the identity matrix and calculate the affinity matrix among the support audio embeddings for the training. The weights of the linear layer are updated using the cross-entropy between the predicted logits and the ground truth of these support examples. Finally, we apply the updated weights to infer the labels of the test audio clips.

\vspace{-0.2cm}
\subsection{Relations with existing works}
\label{subsec: relate_with_existing_work}
\noindent \textbf{Relation with TIP-adapter.} The TIP-adapter was proposed for the image domain whereas our Treff adapter is proposed for audio. In addition, our Treff adapter differs from the TIP-adapter on how the trainable parameters are finetuned. While the TIP-adapter applied a cache model to learn key values in the few-shot knowledge retrieval, our Treff adapter trains a linear layer to transform both query and support audio embeddings to a new space where audio embeddings belonging to the same class get closer than those in the original space. Therefore, the TIP-adapter will lose the information from the support embeddings due to the gradient-optimization while our Treff adapter preserves the semantic information of the support examples during the training. The experimental results in Table~\ref{tab:few_shot} showed that the information retained is beneficial, especially when the computational resource is limited. Meanwhile, when initialising the learnable parameters with appropriate methods and fixing them in the training, both our Treff adapter and TIP-adapter reduce to a cosine similarity measure.

\noindent \textbf{Relation with Matching Network.} While both Matching Networks and Our Treff adapter attempt to retrieve query labels with the affinity matrix between query and support embedding, the matching network~\cite{vinyals_matching_2016} aims to train a generic audio encoder from scratch. Therefore, matching network keeps the affinity calculation fixed throughout the training process. In contrast, benefiting from pretrained audio-language models, the Treff adapter optimises the affinity measurement in the finetuning. This makes the Treff adapter yield a better performance than the matching network with less training resources.  

\noindent \textbf{Relation with Siamese network.} Siamese networks~\cite{koch_siamese_2015} attempted to find a gradient-optimised distance measure between two input examples. Our Treff adapter, however, transforms the embedding to a semantic space where the clusters of data points become more compact. In addition, the Siamese network learns implicitly the feature representation for which our proposed methods leverage a pretrained model. 

%
%
\section{Experiments and results}
\label{sec: experiments_and_results}

\subsection{Experiment setup}
\label{subsec: experiment_setip}
We compare our methods with several sound datasets, including ESC-50~\cite{piczak_esc_2015}, FSDKaggle2018~\cite{fonseca_general-purpose_2018}, and FSD-FS~\cite{liang_learning_2022}. Among them, FSD-FS is a dedicated subset of FSD50K~\cite{fonseca_fsd50k_2022} for multi-label few-shot audio classification.

To implement experiments, we resampled the audio recordings at 44.1kHz. The window length was about 20ms with 50\% overlap, and the number of Mel bank filters was fixed to 64. Log-Mel spectrograms were used as input for few-shot learning methods. We followed the few-shot settings as per~\cite{liang_leveraging_2022,liang_learning_2022}. More details can be found in our released code~\footnote{To be released after peer review}.

\subsection{Experiment results}
\label{subsec: experiment_results}

\vspace{-0.25cm}
\begin{table}[th]
\caption{\small{\textbf{Accuracy (\%) of different methods for the ESC-50 and FSDKaggle18K datasets.} ZS denotes zero-shot performance, and FT is short for few-shot performance. * indicates our reproduced results. The best result on each dataset is marked with \underline{underline}, and the best adaptation methods are highlighted in \textbf{bold}.}}
\vspace{-0.25cm}
\label{tab:fullsize}
\centering
\begin{tabular}{@{}lcc@{}}
\toprule
Model                  & ESC-50         & FSDKaggle18K \\ \midrule
\multicolumn{3}{c}{Topline systems}                     \\
AudioCLIP (FT)~\cite{guzhov_audioclip_2022}         & \underline{97.15}              &         -       \\
CLAP (FT)~\cite{elizalde_clap_2022}              & 95.30*/96.70      &         -       \\
HTS-AT~\cite{chen_hts-at_2022}                 & 97.00              &         -       \\ \midrule
\multicolumn{3}{c}{Baseline systems}                    \\
VGGish (ZS)~\cite{hershey_cnn_2017}            & 33.00              &         -       \\
AudioCLIP (ZS)~\cite{guzhov_audioclip_2022}         & 69.40              &         -       \\
CLAP (ZS)~\cite{elizalde_clap_2022}              & 82.60              &         -       \\ \midrule
Treff adapter (w/o FT) & 87.75              &       71.03   \\
TIP-adapter  (w/ FT) \cite{zhang_tip-adapter_2022}   & 91.45*              &       83.23*   \\
Treff adapter (w/ FT)  & \textbf{92.21}     &\underline{\textbf{86.58}} \\ \bottomrule
\end{tabular}
\end{table}

\textbf{Comparison with fine-tuning.} Table~\ref{tab:fullsize} shows the comparison between the proposed Treff adapter against the state-of-the-art following both fine-tuning (topline systems) and the zero-shot transferring (baseline systems) approach. We show the results of CLAP adapters trained with 16 examples here, as the improvement is getting marginal when the amount of support data increases (see more details in Figure \ref{fig:num_shots_increase}). Our Treff adapter outperforms the zero-shot learning methods by a large margin and even achieves a comparable performance with the fine-tuning methods. In addition, the proposed Treff adapter outperforms the TIP-adapter, which was also adopted as a CLAP adapter, by 0.8\% absolute in terms of accuracy. This is likely because our Treff adapter benefits from the preserved semantic information in the support embedding whereas the TIP-adapter discards it via gradient-based optimisation.

\begin{table}[th]
\centering
\caption{\small{\textbf{Accuracy (\%) of various methods in the few-shot settings.} The best results are highlighted in \textbf{bold}.}}
\vspace{-0.25cm}
\label{tab:few_shot}
\resizebox{\columnwidth}{!}{
\begin{tabular}{@{}lccc@{}}
\toprule
              & \multicolumn{2}{c}{ESC-50 (\%)} & FSD-FS (\%)        \\ \midrule
              & 5-way         & 12-way         & 15-way             \\ \cmidrule(l){2-4} 
ProtoNet~\cite{snell_prototypical_2017}      & 88.18         & 77.70          & 33.02              \\
MatchNet~\cite{vinyals_matching_2016}      & 86.83         & 71.81          & -                  \\
HPN~\cite{liang_leveraging_2022}           & 88.65         & 78.65          & -                  \\
Treff adapter (w/o FT) & 97.49 & 94.68 &                       68.34              \\
TIP-adapter~\cite{zhang_tip-adapter_2022}   & 97.52         & 95.58          & 69.45              \\
Treff adapter & \textbf{98.53}& \textbf{96.29} & \textbf{70.59}     \\ \bottomrule
\end{tabular}}
\vspace{-0.2cm}
\end{table}

\textbf{Comparison with few-shot learning.} Table \ref{tab:few_shot} shows the accuracy of the proposed Treff adapter against the metric-based few-shot learning and the TIP-adapter. Both our Treff adapter and the TIP-adapter achieve better results than other metric-based learning 
algorithms by a large margin. The reason may be that these two adapters utilise a large-scale pretrained model as their encoder. Our proposed Treff adapter outperforms the TIP-adapter by 0.71 percentage points in terms of accuracy, indicating that CALM learns task-specific knowledge while preserving the knowledge from CLAP. Table \ref{tab:compare_xmodality} compares the proposed Treff-adapter with other cross-modality few-shot methods on the ImageNet-ESC~\cite{lin_multimodality_2023}. It can be observed that the Treff adapter achieves the best performance. It outperforms the cross-modality few-shot learning by a large margin as it is able to make use of zero-shot knowledge transferring explicitly while the cross-modality FSL discards it gradually in the parameter optimisation.


\begin{table}[tbh]
\centering
\caption{\small{\textbf{Accuracy (\%) of cross-modality few-shot learning.} The best results are highlighted in \textbf{bold}. Xmodal FSL denotes cross-modality FSL method.}}
\vspace{-0.25cm}
\label{tab:compare_xmodality}
\begin{tabular}{@{}llccc@{}}
\toprule
Dataset          & Method        & 1-shot         & 2-shot         & 4-shot         \\ \midrule
                 & Xmodal FSL    & 35.70          & 45.90          & 51.60          \\
ImageNet-ESC-19  & TIP-adapter   & \textbf{86.45} & 87.89          & 87.11          \\
                 & Treff adapter & 85.26          & \textbf{88.03} & \textbf{87.63} \\ \midrule
                 & Xmodal FSL    & 35.00          & 43.50          & 48.50          \\
ImageNet-ESC--27 & TIP-adapter   & \textbf{86.39} & \textbf{87.31} & 87.03          \\
                 & Treff adapter & 86.02          & 87.04          & \textbf{89.35} \\ \bottomrule
\end{tabular}
\end{table}
\vspace{-0.5cm}
%
%
\section{Conclusion}
\label{sec:conclusion}
We presented the Treff adapter, a training-efficient adapter for CLAP, to boost zero-shot classification performance by making use of a small set of labelled data. Specifically, we designed CALM to retrieve the probability distribution of text-audio clips over classes using a set of audio-label pairs and combined it with CLAP's zero-shot classification results. Furthermore, we designed a training-free version of the Treff adapter by using CALM as a cosine similarity measure. Experiments showed that the proposed Treff adapter is comparable and even better than fully-supervised methods and adaptation methods in low-shot and data-abundant scenarios. While the Treff adapter shows that combining large-scale pretraining and rapid learning of domain-specific knowledge is non-trivial for obtaining generic representations for few-shot learning, it is still limited to audio classification tasks. In the future, we will explore how to use audio-language models in diverse audio domains.
%
%
\section{Acknowledgments}
\label{sec:acknowledge}
This work was supported by the Engineering and Physical Sciences Research Council [grant number EP/T518086/1]. The research utilised Queen Mary's Apocrita HPC facility, supported by QMUL Research-IT, http://doi.org/10.5281/zenodo.438045.

\bibliographystyle{IEEEtran}
\bibliography{refs}

\end{document}